\documentclass{ws-p8-50x6-00}
\begin{document}

\title{Electromagnetic Response Functions of Few--Nucleon Systems}

\author{V.D. Efros$^{1)\,2)}$,  W. Leidemann$^{1)\,3)}$, G. Orlandini$^{1)\,3)}$\\
and E.L. Tomusiak$^{4)}$}

\address{$^{1)}$Dipartimento di Fisica, Universit\'a di Trento\\I--38050 
Povo, Italy\\$^{2)}$Russian Research Centre "Kurchatov Institute"\\
1213182 Moscow, Russia\\$^{3)}$Istituto Nazionale di Fisica Nucleare, Gruppo
collegato di Trento, Italy\\$^{4)}$Department of Physics and Engineering Physics\\ 
and Saskatchewan Accelerator Laboratory,\\
University of Saskatchewan, Saskatoon, Canada S7N 0W0
}


\maketitle

\abstracts{Inclusive electromagnetic reactions in few--nucleon
systems are studied basing on accurate three-- and four--body
calculations. The longitudinal $^4$He$(e,e')$ 
response function obtained at $q\le 600$ MeV/c completely agrees with
experiment. The exact $^4$He spectral function  obtained
in a semirealistic potential model is presented, and the accuracy
of the quasielastic response calculated with its help is assessed,
as well as the accuracy of some simpler approximations for the 
response. The photodisintegration cross section of $^3$He 
obtained with the realistic AV14 NN force plus UrbanaVIII NNN force
agrees with experiment. It is shown that this cross section is very
sensitive to underlying nuclear dynamics in the $E_\gamma\simeq 100$ MeV
region. In particular, the NNN nuclear force clearly manifests itself
in this region.}

\section{Introduction}

Our recent results on the $^4$He$(e,e')$ process and the photodisintegration
cross sections of trinucleons are presented in the talk. The issues of
interest are the accuracy of the conventional 4N formulation for the
$^4$He$(e,e')$ reaction and the accuracy of its description in the 
framework of the spectral function approximation, the sensitivity 
of the photodisintegration cross section to the choice of NN interaction,
and the manifestation of NNN force in the process. The quantities we
study are response functions having e.g. in the $(e,e')$ case the form 
\begin{equation}
R(q,\omega)=\bar{\sum}_{M_0}
\int df|\langle\Psi_f|\hat{O}({\bf q})|\Psi_0(M_0)\rangle|^2
\delta[\omega-(E_f+q^2/(2Am)-E_0)],\label{resp}
\end{equation}
$q$ and $\omega$ being momentum and energy transferred to the nucleus.
We avoid calculating the complete set of final states $\Psi_f$, and we 
obtain the response functions with the method 
explained in refs. \cite{ELO,ELO99} requiring only bound--state type 
calculations. Within the adopted formulation of the
problem the numerical results below are accurate at the per cent 
level. 

\section{The $(e,e')$ response and spectral
function of $^4$He}

In the present report, we present the 
results on the $^4$He$(e,e')$ response in the $q$
range of 300--600 MeV/c. Additional information can be found 
in \cite{ELO99,ELO97a,ELOtr97,ELO98}. 
A non--relativistic 4N
Hamiltonian with central NN potentials reproducing the 
$^1S_0$ and $^3S_1$ NN phase shifts up to high energy has been used 
as a dynamic input. This Trento (TN) potential is listed in Ref. \cite{ELO99}. 
The conventional single--particle transition
operator that includes the Sachs form factors $G_{E}^{p,n}$ with the 
usual relativistic correction has 
been adopted. No further approximations have been used, and
 the FSI is thus fully taken
into account. In Fig. 1 the calculated longitudinal response 
function $R_L(q,\omega)$ is shown with solid curves for 
four values of the transferred momentum $q$ $[{\rm MeV/c}]$ along with 
the Bates \cite{bates} and Saclay \cite{saclay} data. 
(The results for $q\le 500$ MeV/c
differ a bit from those in Refs. \cite{ELO97a,ELOtr97} due to minor
corrections in the formulation, see  \cite{ELO98}.)
One observes a very good agreement with experiment in the
whole range of $q$ values.
(A drop in the data at $q=500$ MeV/c
in the vicinity of $\omega=$250 MeV  
is at variance with the sum rule systematics.)  
\begin{figure}[h]
\vspace{7.8truecm}
\includegraphics{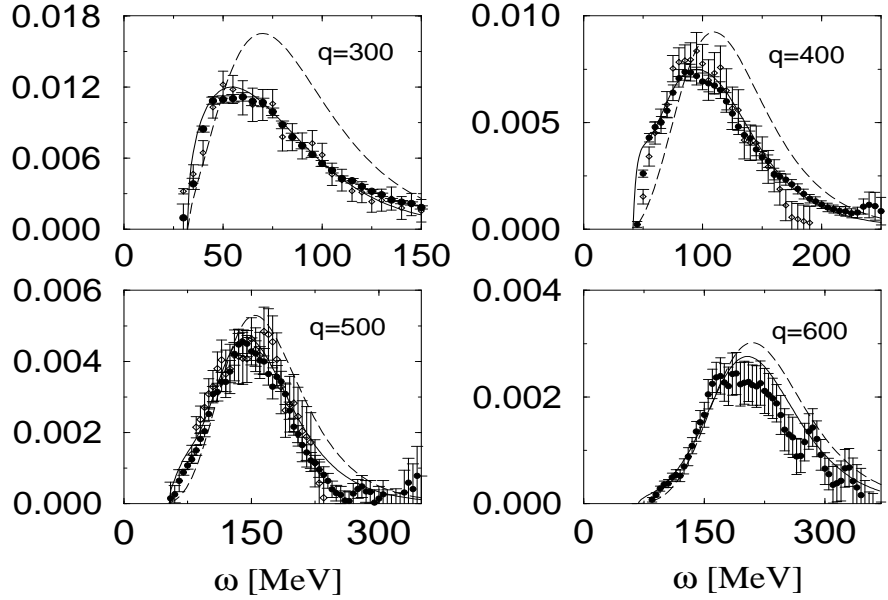}
\caption{Longitudinal response of $^4$He in MeV$^{-1}$. Full results: 
solid curves, PWIA calculation:
dashed curves. Bates and Saclay data are represented with open diamonds
and closed circles, respectively.}
\end{figure}

To interpret the results obtained let us estimate the maximal momenta in 
the initial state contributing substantially to the response. 
Let us 
consider e.g. the final state component in which only the N--$(A-1)$
relative motion is fast. It is sufficient here to consider this relative
motion in the plane wave approximation. 
Let ${\bf k}_{rel}$ be
the relative momentum between the fast nucleon and the remaining $(A-1)$
subsystem.  The FSI 
admixes the momenta lying in the range ${\bf k}_{rel}+\Delta{\bf k}$
to ${\bf k}_{rel}$. 
The momenta $\Delta{\bf k}$ are determined by the nuclear force,
and they are lower than ${\bf k}_{rel}$ values of interest for us. We have
${\bf k}_{rel}=[(A-1)/A][{\bf k}_N-(A-1)^{-1}{\bf K}_{A-1}]=
{\bf k}_N-A^{-1}{\bf K}_A$. 
The corresponding single--particle 
momentum probed in the ground state is
${\bf k}_{miss}\equiv{\bf k}_{rel}-{\bf q}(A-1)/A$. 
The predominant contribution comes
from the states with ${\bf k}_{rel}$ directed along ${\bf q}$. Using the 
energy conservation
in the form $\omega\simeq k^2_{rel}A/[2(A-1)m]+q^2/(2Am)-E_{gs}$
and setting $\omega+E_{gs}=q^2/(2m)+\Delta$ one obtains 
${k_{miss}\simeq[(A-1)/A]|\{q^2+[A/(A-1)]2m\Delta\}^{1/2}-q|}$. Let us
consider the right wing of the response. Setting, for example, 
$q=600$ MeV/c, $\Delta =$100 MeV, c.f. Fig. 1, one gets $k_{miss}=0.7$ fm$^{-1}$. As a consequence, only low momenta in the nuclear interaction
play a role in the kinematical region considered. Contributions from other
reaction mechanisms cannot change this conclusion.
\begin{figure}[h]
\vspace{3.0truecm}
\includegraphics{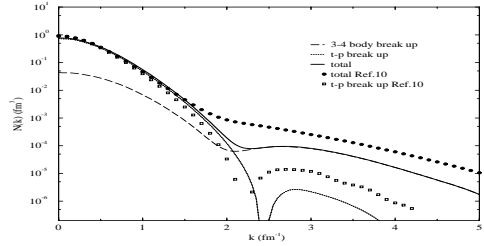}
\caption{Total and partial momentum distributions of $^4$He with the TN potential
and with Argonne $v_{18}+$Urbana IX $^{10}$. }
\end{figure}
The semirealistic NN 
potentials of the type we used probably lead to the same 
low momentum content of
the wave functions as the fully realistic NN interactions. This is illustrated in
Fig. 2 where the single--particle momentum distribution of $^4$He for our NN
potential is compared with that obtained from a realistic nuclear force
\cite{arr95}.
The agreement above with experiment at the right 
wing of the response
then testifies to a correct description of the corresponding low--momentum content of
the wave functions within the adopted 4N formulation. It shows   
the applicability of the conventional form of the electromagnetic 
interaction as well. On the contrary, rather high nucleon momenta in the 
ground--state wave function are probed at the left wing of the response
at high $q$ and small $\omega$ \cite{cg}. 

The SF of $^4$He and the corresponding QE response
have also been calculated exactly with the same NN potential. The SF
$S(k,E)$ represents the 
joint
probability of finding a nucleon with momentum ${\bf k}$ and a residual $(A-1)$
system with energy $E+E_{gs}$:
\begin{equation}
S_{t_{z}}(k,E)=\bar{\sum}_{M_{0},s_{z}}\int df
|\langle\Psi_f^{A-1};{\bf k}s_zt_z|\Psi_0^A(M_0)\rangle|^2
\delta[E-(E_f^{A-1}-E_0^A)].
\label{spf}
\end{equation}
Here $s_z$ and $t_z$ are the spin and isospin quantum numbers of the nucleon,
$E_f^{A-1}$ and $\Psi_f^{A-1}$ are eigenvalues and eigenstates of the $(A-1)$
system, and $E_0^{A}$ and $\Psi_0^{A}$ are such quantities for the ground
state.
The energy $E$ can be viewed as the removal 
energy of the nucleon. The calculation of the SF from its definition 
requires obtaining the complete set of states of the residual $(A-1)$ subsystem. 
Previously the spectral function has been calculated
in the A=3 case, which is essentially the two--body problem, and, 
approximately, for nuclear
matter, see ~\cite{cs} for the references. We have avoided calculating the
complete set of $A=3$ continuum states with the help of the approach mentioned 
in Sec. 1.  
\begin{figure}[h]
\vspace{3.0truecm}
\includegraphics{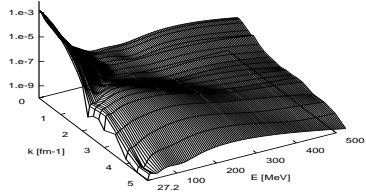}
\caption{SF $S_p(k,E)$ of $^4$He with TN potential in units of fm$^3$ MeV$^{-1}$.}
\end{figure}
The proton SF $S_p(k,E)$ thus obtained is shown in Fig. 3. It includes
two contributions: from the $A=3$ rest subsystem in the ground state,
and in the continuum, c.f. Eq. (\ref{app}) below. The first of these  is determined
by the corresponding momentum 
distribution, and the second 
one is shown in the figure.
The values of 
$S_p(k,E)\ge E_{thr}^{A-1}+1$ MeV are plotted. (We note that 
$S_p(k,E_{thr}^{A-1})=0$ and thus $S_p(k,E)$ exhibits a strong slope at
low energy.) For momenta below 2 fm$^{-1}$ one finds a sharp maximum at
about 2 MeV above $E_{thr}^{A-1}$. 
In Fig. 4 the cut of the calculated spectral function at $k=0.25$ fm$^{-1}$ is shown.
These results appear to have subsequently been confirmed experimentally
 \cite{?}.
\begin{figure}[h]
\vspace{3.0truecm}
\includegraphics{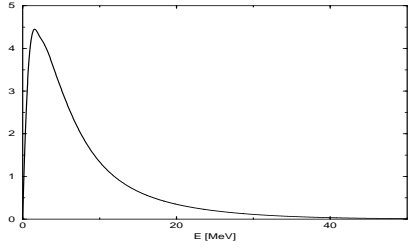}
\caption{The cut of the $^4$He SF at $k=0.25$ fm$^{-1}$.
The values of SF are in 10$^{-3}\cdot$ fm$^3$ MeV$^{-1}$.}
\end{figure}
For $k>2$ fm$^{-1}$ the  calculated spectral function exhibits  a ridge 
where the peak position shifts to higher $E$ for increasing $k$. This 
ridge  should presumably correspond to 
a proton 
struck from the region of a strong two--nucleon correlation, see \cite{cs,fs}. 

The QE response  
\begin{eqnarray}
R^{qe}_L(q,\omega)= A[\tilde{G}_E^p(Q)]^2\int d{\bf k}dE S_p(k,E)\nonumber\\
\times\delta[\omega-E-({\bf k+q})^2/(2m)-k^2/[2(A-1)m]].\label{qe}
\end{eqnarray}
is shown in Fig. 1 with the dashed curve. 
At $q=600$ MeV/c the height of the peak is overestimated by about 9\%
in the framework of the QE approximation.
At low $\omega$, the PWIA assumption inherent to $R^{qe}_L$ obviously breaks 
down. For 
high  $\omega$, the relative 
role of disregarded lower momenta admixtures
in the final state wave function increases with $\omega$.
Due to these reasons the relative error of the QE approximation (\ref{qe}) 
increases at the wings of the response. It is seen from Fig. 1 that the QE peak is shifted to higher 
energies with respect to the true one. The shift of the peak can be qualitatively
understood considering a nucleon at rest in a potential well: $\omega$ can be
estimated as $q^2/(2m)+V_f(q)-V_i$, where $V_{f,i}$ are the potential energies
before or after interaction with the virtual photon. While $V_f$ is negative, it
becomes zero in PWIA leading to an increase in $\omega$.

It is advantageous to have a simple and good approximation for the QE response.
One can write
\begin{equation}
S(k,E)=n_{tp}(k)\delta[E-E_0(^3{\rm H})+E_0(^4{\rm He})]+S_{inel}(k,E).
\label{app}
\end{equation} 
Here the first term represents the residual $A=3$ system in the ground state, 
and we do not differentiate between the t-p and n-$^3$He channels. The 
second term is depicted in Fig. 3 and 
represents the contribution of the residual $A=3$ system being in 
continuum. It is seen from Fig. 3
that for low $k$ relevant to us almost all the strength in $S_{inel}$
is concentrated close to the thershold $E=E_{min}$. This suggests the
approximation (to be put into Eq. (\ref{qe}))
\begin{equation}
S(k,E)\simeq n_{tp}(k)\delta[E-E_0(^3{\rm H})+E_0(^4{\rm He})]+
n_{t^*p}(k)\delta[E-E_{t^*p}+E_0(^4{\rm He})]\label{app1}
\end{equation} 
where $E_{t^*p}$ is the threshold breakup energy of the residual nucleus. 
Using (\ref{app1}) 
one does not need continuum state calculations since
$n_{t^*p}(k)=n(k)-n_{tp}(k)$. One obtains an even simpler approximation
considering that $E_{t^*p}\simeq E_0(^3{\rm H})$:
\begin{equation} 
S(k,E)\simeq n(k)\delta[E-E_0(^3{\rm H})+E_0(^4{\rm He})].\label{app2}
\end{equation} 
Eq. (\ref{app2}) was used in the literature (see e.g. \cite{cas}). 
One may note that at the small $k$ values of interest the term $S_{inel}$ in
(\ref{app}) is only of secondary importance (e.g. the integrated values of
$n_{tp}$ and  $n_{t^*p}$ equal 0.89 and 0.11, respectively,) which improves 
the quality of the approximations (\ref{app1}), (\ref{app2}). 
\begin{figure}[h]
\vspace{3.0truecm}
\includegraphics{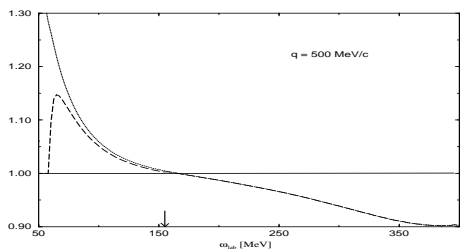}
\caption{QE response (\ref{qe}) with the SF of Eq. (\ref{app1}) (dashed curve)
 and of Eq. (\ref{app2}) (dotted curve) relative to that with the full SF (the QE 
peak is marked by an arrow).}
\end{figure}
In Fig. 5 we show the responses (\ref{qe}) obtained
with these approximations relative
to Eq. (\ref{qe}) with the full spectral function. Except for  
small $\omega$
where the wrong threshold behavior in Eq. (\ref{app2}) may play a role,
the three responses are very similar, particularly in the peak region. (At $q=$
300, 400, and 1000 MeV/c one has very similar results.) 

\section{Total photodisintegration cross sections of trinucleons}

Earlier theoretical work  on the 
photodisintegration cross section of trinucleons in a wide 
energy range  is outlined in Ref. \cite{ELO97b}. The first
rigorous calculations have been performed with separable central
potentials. Mixed symmetry states have been disregarded even though they play
an important role in the process. Approximate calculations 
of the three--body breakup
with low--energy central local potentials were also performed. 
In \cite{vostr} the cross section has been obtained with a non--central
potential but only the three lowest hyperspherical terms in the expansion of 
continuum wave functions were retained which may influence the 
results. In \cite{sch97} the $^3$H two--body photodisintegration 
differential cross sections 
at $\theta=90^o$ were
calculated with Paris and Bonn NN realistic potentials for $E_\gamma\le 40$ MeV. 
In \cite{ELO97b} accurate calculations of the total cross sections have been performed
with central TN and Malfliet--Tjon (MT) \cite{mt} potentials reproducing the 
$s$--wave phase shifts up to high energy. 

In the present work we 
calculate the total photodisintegration cross section with realistic NN interactions 
for the first time and we include an NNN force in addition. 
As in \cite{ELO97b}, the approach outlined in Sec. 1 has been used. 
The calculations are performed 
for photon
energies up to pion threshold, and we pay special attention to 
the high energy tail
of the cross section. We obtain a sizeable contribution of NNN force 
to the cross section and also comment on the sensitivity to the NN force.
It may be natural to expect some sensitivity
to the NN force since
if one takes, for example, $E_\gamma= $110 MeV and considers the 
N$+d$ breakup then the N$-d$ relative momentum is 
about 1.9 fm$^{-1}$, and  single--particle 
momenta in $\hat {O}\Psi_0$ of this size 
may be probed in the process. 

The calculation was done with a non--relativistic three--nucleon Hamiltonian. 
The realistic NN interactions AV14 \cite{av14} and TRSB \cite{trs}
were employed. These NN interactions
underbind $^3$H by about 0.8 MeV and 1 MeV, respectively. In the third version
of the calculation, the AV14 NN interaction plus NNN UrbanaVIII force are used. 
Along with these results  the cross sections \cite{ELO97b} 
with central TN and MT potentials are listed as well.

For the latter potentials, as well as in the case of the softer core
TRSB interaction,
 we used the Jastrow correlated hyperspherical basis \cite{fe72}. 
The correlations speed up the convergence but not to a sufficient degree in case
of the stronger core AV14 interaction. In the Jastrow case, one chooses
the two--body correlation function as the mean value of the triplet and singlet
NN low--energy wave functions. However, for interactions
like AV14 it is advantageous to take 
into account differences between correlations
in different two--nucleon spin--isospin states. One possibility is to use
different Jastrow factors at different groups of basis states before
symmetrization \cite{suki}. However, after symmetrization the correlations 
then lose their
role as correlations in given NN states and become some effective 
correlations which still may substantially speed up the convergence \cite{suki} 
due to the
many variational parameters involved. Such a procedure is not convenient in our 
case since our response calculations are not variational ones. In the present work, in the AV14 case,
we use a new version  of a correlated basis. Spin--isospin 
dependent correlations are used, and they
retain their role as correlations in the given four spin--isospin NN states, see 
~\cite{leid99}. 
In Table 1 the properties of our
ground state wave function of $^3$H 
are listed in comparison with Ref. \cite{chen}. The 
UrbanaVIII potential was taken as the NNN force in the last column.

\begin{table}[t]
\caption{Bound--state properties of $^3$H with the AV14 interaction}
\begin{center}
\footnotesize
\begin{tabular}{|c|c|c|c|c|}
\hline
 & $E_b$, MeV& $  r_{rms}$, fm &$P(D)$, \% & $E_b$(NN+NNN), MeV\\
\hline
present work & 7.71&1.66 &8.85 & 8.52\\
\hline
Ref. \cite{chen} &7.67&1.67 &8.96 & 8.46\\ 
\hline
\end{tabular}
\end{center}
\end{table}

In the present work we confine ourselves with the unretarded E1 approximation
for the transition operator which is known to be a very good one for
the total cross section at least for not too high energies. Within this approximation
the transition current operator may be rewritten exactly as the dipole moment
operator of a system,
while MEC currents are taken into  account automatically.
\begin{figure}[h]
\vspace{3.0truecm}
\includegraphics{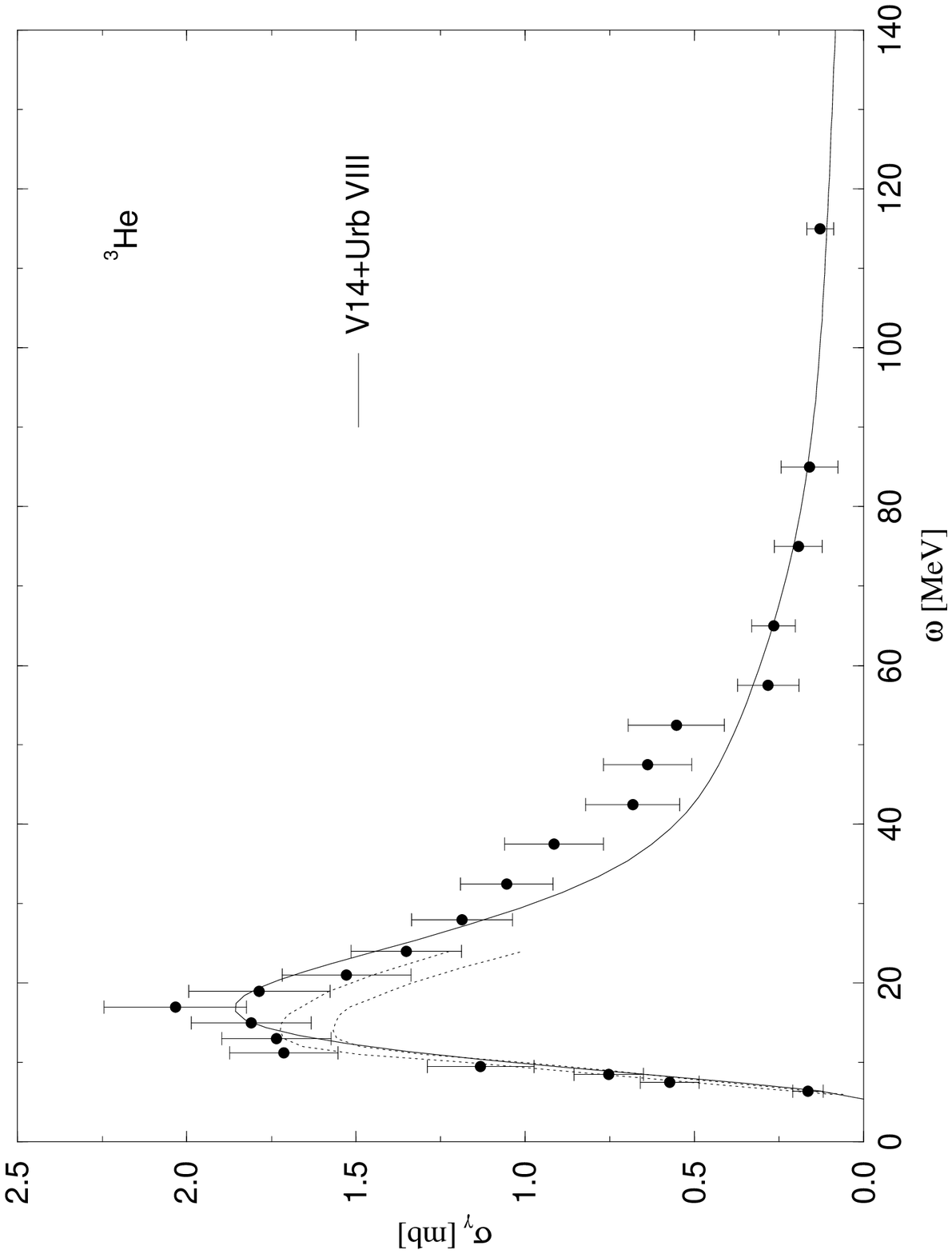}
\caption{Photodisintegration cross section of $^3$He. Theoretical curve
is obtained with the 
AV14+UrbanaVIII nuclear force (full curve). Experimental data are of 
Refs. $^{24}$ 
(filled circles) and $^{25}$ (upper 
and lower bound delineated with dotted curves).}
\end{figure}
In Fig. 6 our total photodisintegration cross section of $^3$He calculated with
AV14+UrbanaVIII interaction is shown 
along with experimental data. The theory leads to an agreement with
the data of Ref. \cite{fet}. In Fig. 7 our results 
in the tail region
of the $^3$H cross section are shown. 
Separate cross sections for transitions to final states with 
isospin $T=1/2$ and 3/2 are presented. 
Transitions to $T=3/2$ correspond to the three--body breakup while those to
$T=1/2$ are known to correspond predominantly to the two--body breakup. \begin{figure}[h]
\vspace{3.0truecm}
\includegraphics{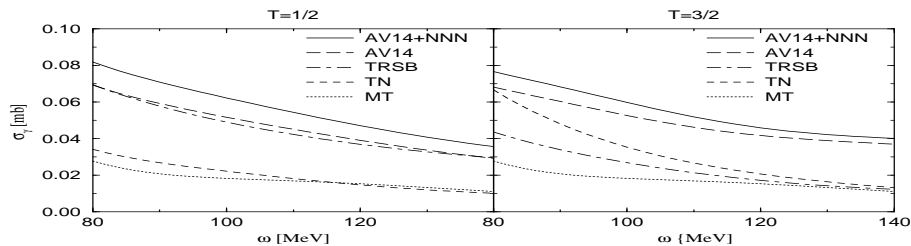}
\caption{Photodisintegration of $^3$H. Results with various choices of nuclear
interactions. The UrbanaVIII potential was taken as the NNN force.} 
\end{figure}
One sees a large difference between predictions of different models in this
region. In the $T=1/2$ case, realistic NN interactions lead to much higher 
cross section than central NN forces. This can be attributed to tensor correlations
produced by realistic interactions. It is also seen that the NNN nuclear force
clearly manifests itself by leading to a sizable further increase in the cross section.
As to the 
$T=3/2$ case, the interesting point is that the  
cross section obtained with 
the realistic TRSB NN interaction
 is close to those for the
central NN potential models but strongly differs from that obtained with AV14. 
We are planning to consider other realistic NN interactions as well 
to clear out a sensitivity to the choice of NN interaction. 

There exist only two experimental data points in the energy range 
presented in Fig. 7
(see Fig. 6, the $^3$He and $^3$H cross sections
 are practically indistinguishable
in this region \cite{ELO97b}). These experimental data are in 
favour of the AV14+UrbanaVIII interaction. Our results thus 
show that, in contrast 
to many quantities conventionally studied in few--nucleon physics,  
an experimental
study of the process in the considered energy region 
may allow a clear discrimination between various models of NN
interaction and detection of the contribution of NNN force.
Corrections from retardation and the contribution of other 
multipoles will be studied separately. They are rather small,
and they cannot change the
conclusion about high sensitivity of the process to nuclear dynamics.
Obviously, one 
can anticipate even more sensitivity to nuclear dynamics in exclusive experiment.

\end{document}